\newcommand{\beqn}{\begin{equation}}
\newcommand{\eeqn}{\end{equation}}

\documentclass[aps, pra, reprint, groupedaddress]{revtex4-2}

\usepackage{dcolumn}
\usepackage{bm}

\usepackage{epsfig}
\usepackage{graphicx}
\usepackage{amssymb}
\usepackage{amsmath}

\begin{document}

\title{Diode laser overtone spectroscopy of methane at 780 nm}

\author{A. Lucchesini}
\email{lucchesini@ino.cnr.it}
\affiliation{Istituto Nazionale di Ottica - CNR - S.S. ``Adriano Gozzini'' \\
Area della Ricerca - Pisa - Italy}

\date{\today}


\begin{abstract}
A tunable diode laser spectrometer has been employed to examine the overtone absorption lines of $^{12}$CH$_4$ in the region around 12780 cm$^{-1}$ (780 nm) at room temperature. The spectrometer sources are commercially available tunable diode lasers operating in ``free-running'' mode. The measurement of the line positions within 0.01 cm$^{-1}$ has been possible by the use of the wavelength modulation spectroscopy and the second harmonic detection techniques that permitted the observation of minimum optical absorbances of the order of $\approx 10^{-6}$. The weakest observed lines of the band have absorption cross sections of the order of $1 \times 10^{-25}$ cm$^2$/molecule, that is $\simeq 0.3$ km$^{-1}$/amagat. For some of them self-, air-, He- and H$_2$-broadening coefficients have been obtained.
\end{abstract}
\maketitle



\section{Introduction}
$^{12}$CH$_4$ is an extremely interesting molecule present in the spectra of the outer planets of the solar system like Uranus, Neptune, Jupiter, Saturn and the relative satellites, in particular Titan~\cite{ref:kui}, where it is the second most abundant gas in the atmosphere; both methane and ethane have also been found in the Kuiper Belt~\cite{ref:bro}.
In spite of their weakness, overtone and combination tone absorption lines of the methane gas are observable in the absorbing layers of the planets~\cite{ref:kark} because these are much thicker than the ones obtainable in the laboratory. The knowledge of the methane optical resonances and their behavior with pressure is important for a better knowledge of the planetary atmospheres.
Conventional spectroscopy measurements of this {\it spherical top molecule} in the visible and in the near infrared have been systematically performed in the past by analyzing the telescope photographic plates~\cite{ref:vedder} or by using long path White cells in the laboratories~\cite{ref:giver,ref:dick}, but to detect the weak overtone bands, advanced technologies like the intracavity laser spectroscopy (ILS) have been used~\cite{ref:singh,ref:brien}.\\
After the previous work on the $\nu_1 + 3\nu_3$ methane absorption band~\cite{ref:alex} another spectroscopic work based on the use of the semiconductor diode lasers (DLs) is presented here. The frequency modulation (FM) technique comes to help when dealing with very weak absorption resonances taking advantage of one of the more interesting properties of the DLs, that is the modulability through the injection current. In this work we apply the FM spectroscopy, usually called ``wavelength modulation spectroscopy'' (WMS) when the value of the frequency is chosen much lower than the resonance line-width, and the second harmonic detection techniques to the methane absorptions around 780 nm, where the very weak CH$_4$ combination overtone band $3\nu_1 + \nu_3 + (\nu_2$ or $\nu_4)$ are located.

\section{Experiment}

\subsection{Experimental setup}
\label{exp}

\begin{figure}
\resizebox{\columnwidth}{!}{
  \includegraphics{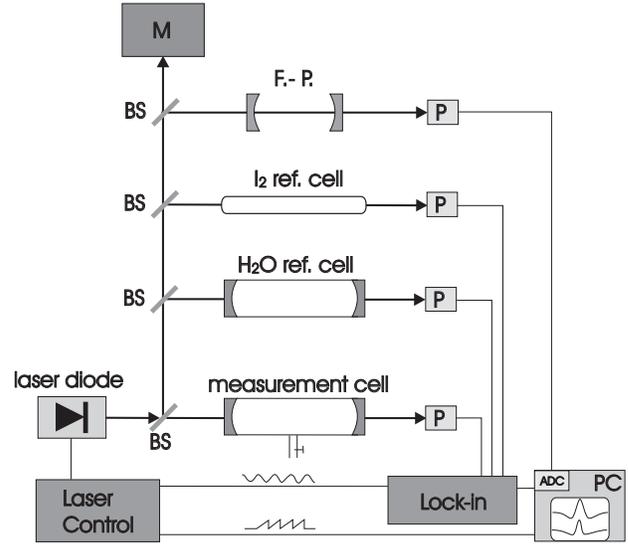}
}
\caption{Schematic drawing of the experimental apparatus. ADC: analog to digital converter; BS: beam splitter; F.-P.: Fabry-Perot interferometer; M: monochromator; P: photodiode; PC: desk-top computer.}
\label{appar}
\end{figure}
Fig.~\ref{appar} shows the experimental setup adopted for the WMS in this work.
The arrangement has been described in our previous work~\cite{ref:alex}. The source is a Fabry--Perot type semiconductor heterostructure AlGaAs laser SHARP LT025MD that emits a 40 mW continuous wave (cw) at 780 nm without any external optical feedback. Thanks to its V-shaped junction cladding layers based on SHARP's original technology (VSIS chip structure) this type of cw diode laser has a single transverse and single longitudinal mode. In this ``free-running'' configuration the full width at half the maximum (FWHM) of its frequency emission mode is around 20 MHz. Its operational current amplitude ranges between 70 and 100 mA. The measurement cell containing the sample gas is a multipass custom made Herriott type, 30 m path length. Another Herriott cell contains water vapor for checking whether the obtained absorption features come from H$_2$O that could contaminate the cell. An iodine reference glass cell is used for the precise wavenumber measurements. For the harmonic detection, a sinusoidal current is mixed to the diode laser injection current and the signals collected by pre-amplified silicon photodiodes are sent to the lock-in amplifiers in order to extract the desired harmonic components. The methane gas was supplied by Praxair Inc. with the following nominal characteristics: grade 5.5 (purity $\geqslant$ 99.9995\%), H$_2$O $\leqslant 2$ ppmv, O$_2 \leqslant 0.5$ ppmv, N$_2 \leqslant 10$ ppmv, H$_2 \leqslant 0.1$ ppmv.

\subsection{Frequency modulation}
\label{freq}
The measured transmittance of the gas samples $\tau(\nu)$ is well described by the Lambert-Beer equation:
\beqn
\tau(\nu) =  e^{-\sigma(\nu) \, z}
\label{eq:beer}
\eeqn
where $z=\rho\,l$ is the column amount (in molecule/cm$^2$), i.e. the product of the absorbing species number density $\rho$ (in molecule/cm$^3$) and the optical path $l$ (in cm) of the radiation through the sample; the absorption cross section $\sigma(\nu)$ is expressed in cm$^2$/molecule. In case of small optical depths [$\sigma(\nu) z \ll 1$], always verified in our case, Eq.~(\ref{eq:beer}) can be approximated by
\beqn
\tau(\nu) \simeq 1 - \sigma(\nu)\, z \, .
\label{eq:absor}
\eeqn
$\sigma(\nu)$ must take into account the shape of the absorption line: Gaussian-like for Doppler, inhomogeneous broadening and Lorentzian-like for collisional, homogeneous broadening. Other effects, like the Dicke narrowing that occurs when the molecular mean free path is comparable to the wavelength of the radiation~\cite{ref:dicke}, are not observed in our measurement conditions and are not taken into account. For the purpose, the Voigt function well describes the behavior of the absorption as a function of the radiation frequency:
\beqn
f(\nu)=\int_{-\infty}^{+\infty}{\frac{\exp{[-(t-\nu_{\circ})^2 /
{\it\Gamma_\mathrm{G}^\mathrm{2}}\ln2]}}{(t-\nu)^2+{\it\Gamma_\mathrm{L}^\mathrm{2}}}}dt
\label{eq:voigt}
\eeqn
where $\nu_{\circ}$ \ is the gas resonance frequency, ${\it\Gamma_\mathrm{G}}$ and ${\it\Gamma_\mathrm{L}}$ are the Gaussian and the Lorentzian half-widths at half the maximum (HWHM) respectively.
We used the FM technique and therefore the emission frequency $\bar{\nu}$ of the DL  was sinusoidally modulated at the frequency $\nu_\mathrm{m} = \omega_\mathrm{m}/2\pi$, resulting in
\beqn \label{eq:freq}
\nu = \bar{\nu} + a \cos \omega_\mathrm{m} t \, .
\eeqn
In this case the transmitted intensity depends on both the line shape and the modulation parameters and can be written as a cosine Fourier series:
\beqn
\tau(\bar{\nu} + a\, \cos \omega_\mathrm{m} t) =
\sum_{n=0}^\infty  H_n(\bar{\nu}, a) \, \cos n \omega_\mathrm{m} t
\label{eq:tau}
\eeqn
where $H_n(\bar{\nu})$ is the $n$-th harmonic component of the modulated signal. By using a lock-in amplifier tuned to a multiple $n \nu_\mathrm{m}$ ($n = 1,2,...$) of the modulation frequency, the output signal is proportional to the $n$-th component $H_n(\bar{\nu})$ and when the amplitude $a$ is chosen smaller than the width of the transition line, the $n$-th Fourier component is proportional to the $n$-order derivative of the original signal:
\beqn
H_n(\bar{\nu}, a) = \frac{2^{1-n}}{n!} \, a^n \,
\left. \frac{d^n \tau(\nu)}{d\nu^n}\right|_{\nu=\bar{\nu}} ,
\quad\quad n \geqslant 1 \, .
\label{eq:comp}
\eeqn
For pressure broadening measurements we used a low modulation amplitude and we detected the second harmonic component (2$f$ detection), therefore the output signal was proportional to the second order derivative of the real absorption line. This enhanced the signal-to-noise (S/N) ratio and reduced to zero the unwanted background. In order to extract the line parameters, we used a nonlinear least-squares fit procedure explained elsewhere~\cite{ref:rosa}. In particular we fitted the full-width at half the maximum (FWHM) $\gamma_\mathrm{L}$ vs pressure by the general expression:
\beqn
\gamma_\mathrm{L}(p) = 2{\it\Gamma_\mathrm{L}}(p) = \gamma_i p_i + \gamma_\mathrm{self} p_{\circ}
\label{eq:broad}
\eeqn
where $\gamma_i$ is the FWHM broadening coefficient related to the $i$ buffer gas, $p_i$ is the buffer gas partial pressure, $\gamma_\mathrm{self}$ is the sample gas FWHM self-broadening coefficient and $p_{\circ}$ is the sample gas partial pressure.

\subsection{High modulation regime}
\label{eq:modul}
\begin{figure}
\resizebox{\columnwidth}{!}{
  \includegraphics{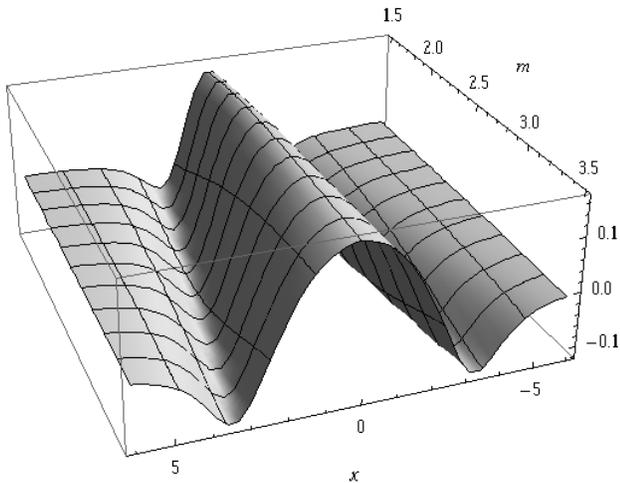}
}
\caption{Behavior of Eq.~(\ref{eq:second}) as a function of the modulation parameter $m$.}
\label{modbrol}
\end{figure}

To obtain the line positions even for the weakest lines it was necessary to use large values of the modulation amplitude. This substantially improves the S/N ratio, but when $a$ is increased, the derivative approximation of Eq.~(\ref{eq:comp}) fails and the $n-$th harmonic component $H_n(\nu,a)$ becomes~\cite{ref:web}
\beqn
H_n(\nu,a) = \frac{2}{\pi} \int_0^\pi \tau(\nu + a \cos \theta)
\cos n\theta \; d\theta \, .
\label{eq:Hn}
\eeqn
The analytical evaluation of this integral is not always possible.
Arndt \cite{ref:arn} and Wahlquist \cite{ref:wahl} derived the analytical form of the harmonic components for a Lorentzian absorption function. The expression for the $n-$th harmonic component can be obtained by inverting Eq.~(\ref{eq:tau}):
\beqn
H_n(x,m) = \varepsilon_n \, \mathrm{i}^n \,
\int_{-\infty}^{+\infty} \hat{\tau}(\omega) \, J_n(m \omega) \,
e^{\mathrm{i} \omega x} \, d\omega
\label{inversa}
\eeqn
where
\beqn
\hat{\tau}(\omega) = \frac{1}{2\pi} \int \tau(x) \, e^{-\mathrm{i} \omega x} \, dx
\eeqn
is the Fourier transform of the transmittance profile; $x=\nu/{\it\Gamma}$ and $m=a/{\it\Gamma}$ are respectively the frequency and the amplitude of the modulation, normalized to the line-width ${\it\Gamma}$; $J_n$ is the $n-$th order Bessel function; $\varepsilon_0=1$, $\varepsilon_n=2$ ($n=1,2,\cdots$) and i is the imaginary unit. Assuming a Lorentzian absorption line-shape centered at $\nu=0$ (this is acceptable when, as in our case, collisional broadening dominates) the cross-section coefficient will be:
\beqn
\sigma_\mathrm{L}(x,m) \propto \frac{1}{1+(x+m \ cos\omega t)^2}
\eeqn
Referring to the work of Arndt we recalculated the second Fourier component of the cross-section coefficient by putting $n=2$:
\beqn
H_2(x,m) = - \frac{1}{m^2}
\biggl[\frac{\{[(1-\mathrm{i}x)^2+m^2]^{1/2}-(1-\mathrm{i}x)\}^2} {[(1-\mathrm{i}x)^2 + m^2]^{1/2}} + \mathrm{c.c.}\biggr]
\label{eq:imagi}
\eeqn
and by eliminating the imaginary part:
\begin{widetext}
\beqn
H_2(x,m) = \frac{2}{m^2} - \frac{2^{1/2}}{m^2} \times \frac{1/2[(M^2+4x^2)^{1/2}+1-x^2][(M^2+4x^2)^{1/2} + M]^{1/2}+|x|\, [(M^2+4x^2)^{1/2}-M]^{1/2}}{(M^2+4x^2)^{1/2}}
\label{eq:second}
\eeqn
\end{widetext}
where $$M=1-x^2+m^2 \, .$$
Fig.~\ref{modbrol} shows the behavior of Eq.~(\ref{eq:second}) as a function of the modulation parameter $m$. It can be seen that it is
proportional to the 2nd derivative of the absorption feature only for low modulation; for $m = 3$ it is completely deformed by broadening, as it happens in the reality.
This function, properly normalized, has been used to fit all the features obtained in this work by the 2$f$ detection spectroscopy with high modulation ($m=2.2-2.3$).

\section{Experimental results}
\label{res}
112 absorption lines have been observed in this work and their position measured within 0.01 cm$^{-1}$ by the comparison with a very precisely known I$_2$ absorption spectrum~\cite{ref:ger} coming from the I$_2$ reference cell. Table~\ref{tab:list} shows the results with the wavenumbers $\nu'$ in vacuum. They presumably belong to the combination overtone band $3\nu_1 + \nu_3 + (\nu_2$ or $\nu_4)$~\cite{ref:giver,ref:herz} or perhaps $2\nu_1 + \nu_2 +  2\nu_3$~\cite{ref:herz}, where $\nu_1$ is the symmetric, $\nu_3$ is the asymmetric stretch and $\nu_2$ is the bending fundamental vibrations. There are a few gaps in the Table ought to the lack of emission wavelengths of the available laser diodes, therefore this Table does not cover all the lines of the band but most of them.
With the derivative spectroscopy method adopted here it is difficult to obtain absolute intensity measurements directly from the 2$f$ signals. We tried to measure the absolute transmission values using the direct absorption (DA) technique by subtracting the background, at room temperature (RT) on the same path-length as the one adopted with WMS and with the CH$_4$ pressure around 90 Torr that gives the best S/N. The maximum absorption cross sections shown in the Tables for most of the lines are obtained in this way.

\begin{table}
\caption{List of the observed CH$_4$ lines along with the maximum absorption cross sections.}
\label{tab:list}
\begin{tabular*}{.48\textwidth}{@{} c @{\extracolsep{\fill}} c c @{\extracolsep{\fill}} c @{}}
\noalign{\smallskip} \hline \noalign{\smallskip}
$\nu'$ (cm$^{-1}$) & $\sigma_\mathrm{max}$ & $\nu'$ (cm$^{-1}$) & $\sigma_\mathrm{max}$ \\
 & ($10^{-24}\frac{\mathrm{cm}^{2}}{\mathrm{molecule}}$) & & ($10^{-24}\frac{\mathrm{cm}^{2}}{\mathrm{molecule}}$) \\
\noalign{\smallskip} \hline \noalign{\smallskip}
12836.84 & $0.14\pm 0.05$ & 12797.29 & $0.28\pm 0.05$ \\
12812.43 &                & 12794.27 & $0.18\pm 0.05$ \\
12812.34 & $0.10\pm 0.05$ & 12794.13 & $0.15\pm 0.05$ \\
12812.14 &                & 12793.62 &                 \\
12811.78 & $0.31\pm 0.05$ & 12793.54 &                 \\
12809.62 &                & 12793.37 & $0.49\pm 0.05$ \\
12809.53 &                & 12793.24 &                 \\
12809.43 & $0.33\pm 0.05$ & 12793.13 &                 \\
12807.80 & $0.67\pm 0.05$ & 12792.43 & $0.30\pm 0.05$ \\
12807.71 &                & 12792.02 &                 \\
12806.52 & $1.06\pm 0.05$ & 12791.87 & $0.31\pm 0.05$ \\
12804.95 & $0.68\pm 0.05$ & 12791.19 & $0.19\pm 0.05$ \\
12804.73 &                & 12790.98 &                 \\
12804.60 & $0.17\pm 0.05$ & 12790.55 &                 \\
12803.50 & $0.26\pm 0.05$ & 12790.47 & $0.30\pm 0.05$ \\
12803.08 & $0.29\pm 0.05$ & 12781.82 & $0.38\pm 0.05$ \\
12801.96 & $0.08\pm 0.03$ & 12781.41 & $0.56\pm 0.05$ \\
12801.84 & $0.48\pm 0.03$ & 12781.11 & $0.31\pm 0.05$ \\
12800.69 & $0.33\pm 0.03$ & 12779.52 & $0.49\pm 0.05$ \\
12799.09 & $0.42\pm 0.05$ & 12779.13 & $0.32\pm 0.05$ \\
12798.98 &                & 12778.58 & $0.34\pm 0.05$ \\
12798.94 &                & 12778.27 & $0.24\pm 0.05$ \\
12798.63 &                & 12777.81 &                 \\
12798.56 & $0.21\pm 0.05$ & 12777.72 &                 \\
12798.39 & $0.18\pm 0.05$ & 12777.65 &                 \\
12798.22 &                & 12777.41 & $0.19\pm 0.05$ \\
12798.10 &                & 12777.34 & \\
12797.98 &                & 12775.61 & $0.15\pm 0.05$ \\
\noalign{\smallskip}
\hline
\end{tabular*}
\end{table}

\begin{figure}
\resizebox{\columnwidth}{!}{
  \includegraphics*{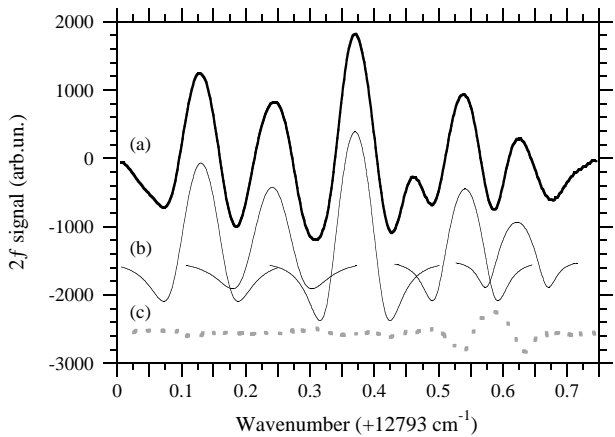}
}
\caption{2nd derivative signal of the methane transmission spectrum around 781.4 nm (a) obtained by WMS with 10 Hz bandwidth at p$_\mathrm{CH_4} = 84$ Torr and T = 296 K. The peaks extracted by the fit procedure (b) are shifted down for clearness. The I$_2$ reference signal (c) is shown dotted.}
\label{spect}
\end{figure}

\setcounter{table}{0} 
\begin{table}
\caption{List of the observed CH$_4$ lines (continued).}
\begin{tabular*}{.48\textwidth}{@{} c @{\extracolsep{\fill}} c c @{\extracolsep{\fill}} c @{}}
\noalign{\smallskip} \hline \noalign{\smallskip}
$\nu'$ (cm$^{-1}$) & $\sigma_\mathrm{max}$ & $\nu'$ (cm$^{-1}$) & $\sigma_\mathrm{max}$ \\
 & ($10^{-24}\frac{\mathrm{cm}^{2}}{\mathrm{molecule}}$) & & ($10^{-24}\frac{\mathrm{cm}^{2}}{\mathrm{molecule}}$) \\
\noalign{\smallskip} \hline \noalign{\smallskip}
12775.42 & $0.16\pm 0.05$ & 12756.65 &                 \\
12775.28 & $0.24\pm 0.05$ & 12756.36 & $1.0\pm 0.1$   \\
12774.93 & $0.14\pm 0.05$ & 12755.48 & \\
12774.72 &                & 12755.36 & \\
12774.27 & $0.45\pm 0.05$ & 12755.13 & $0.41\pm 0.05$ \\
12773.83 & $0.77\pm 0.05$ & 12755.00 & \\
12767.30 &                & 12753.52 & $0.79\pm 0.08$ \\
12767.14 & $0.90\pm 0.06$ & 12753.44 & \\
12767.00 &                & 12752.18 & \\
12762.97 & $0.26\pm 0.05$ & 12752.08 & \\
12762.29 &                & 12751.08 & $1.0\pm 0.1$   \\
12762.20 &                & 12749.46 & $0.25\pm 0.05$ \\
12762.00 & $0.11\pm 0.05$ & 12749.12 & \\
12761.92 &                & 12746.48 & $0.50\pm 0.05$ \\
12761.83 &                & 12746.05 & $0.31\pm 0.05$ \\
12761.33 & $0.14\pm 0.05$ & 12742.99 & $0.33\pm 0.05$ \\
12760.86 & $0.35\pm 0.05$ & 12742.21 & $0.30\pm 0.05$ \\
12760.73 & $0.64\pm 0.05$ & 12741.22 & $0.22\pm 0.05$ \\
12760.57 & $0.22\pm 0.05$ & 12740.15 & $0.39\pm 0.05$ \\
12759.52 &                & 12737.32 & $0.16\pm 0.05$ \\
12759.81 & $0.22\pm 0.05$ & 12729.71 & $0.4\pm 0.1$ \\
12759.42 & $0.4\pm 0.1$   & 12729.43 & $0.17\pm 0.05$ \\
12759.23 &                & 12721.87 & $0.20\pm 0.05$ \\
12757.74 &                & 12714.64 & \\
12757.48 &                & 12714.48 & $0.55\pm 0.06$ \\
12757.19 & $0.94\pm 0.05$ & 12713.30 & $0.92\pm 0.06$ \\
12757.11 &                & 12712.19 & $0.27\pm 0.05$ \\
12756.90 & $0.38\pm 0.05$ & 12695.17 & $0.22\pm 0.06$ \\
\noalign{\smallskip}
\hline
\end{tabular*}
\end{table}
\noindent An example of a crowded methane spectrum collected by using the WMS and the 2nd derivative technique is shown in Fig.~\ref{spect}, where the methane lines are extracted by the fit procedure. Their positions are: 12793.13 cm$^{-1}$, 12793.24 cm$^{-1}$, 12793.37 cm$^{-1}$, 12793.54 cm$^{-1}$ and 12793.62 cm$^{-1}$. The I$_2$ absorption line registered contemporary with methane is shown in dotted line. Its position is known to be $(12793.5780 \pm 0.0017)$ cm$^{-1}$ from the I$_2$ absorption lines database~\cite{ref:ger}. When dealing with free-running diode lasers, a frequency sweep of their emission is always associated with a variation of the emission intensity, therefore in Fig.~\ref{spect} the relative intensity ratios does not give the real ratios as they suffer this intensity change; in particular as the wavenumber increases the intensity of the DL mode decreases. In the range of values of the Figure the DL mode intensity changes by 35\%. In Table~\ref{tab:list} the cross-section is given only for the more intense of them.

\noindent The complexity of the structure of the overtone band does not permit an immediate and correct quantum classification of the ro-vibrational resonances. For these highly excited levels the numerous possible resonances between ro-vibrational levels can modify significantly the intensity as well as the position of the expected lines~\cite{ref:hal}. An aid to the classification job can come from working at very low temperature ($\leqslant 20$ K) using the supersonic methane jet expansion~\cite{ref:per}. In this case only a few first rotational levels will be populated and the level superpositions will be smaller or absent.

\section{Line broadening measurements}
\label{coefficients}
\begin{figure}
\resizebox{\columnwidth}{!}{
  \includegraphics{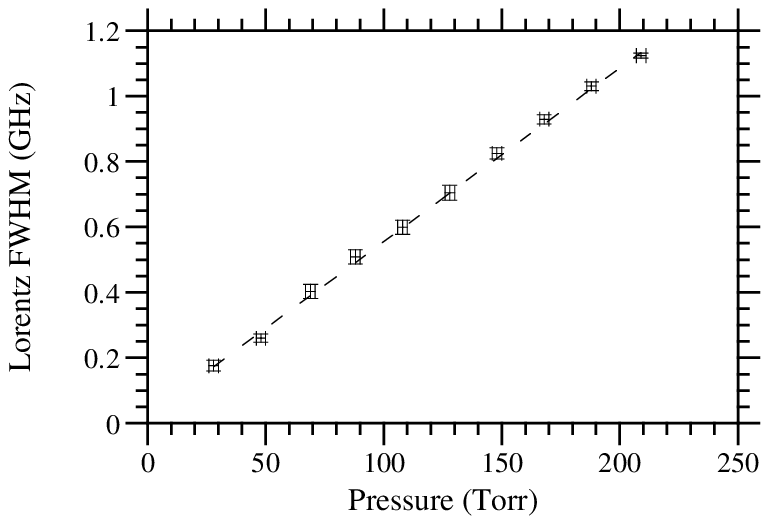}
}
\caption{FWHM self-broadening measurements for the 12713.30 cm$^{-1}$ methane absorption line as a function of pressure  at room temperature.}
\label{ch4self}
\end{figure}

\begin{figure}
\resizebox{\columnwidth}{!}{
  \includegraphics{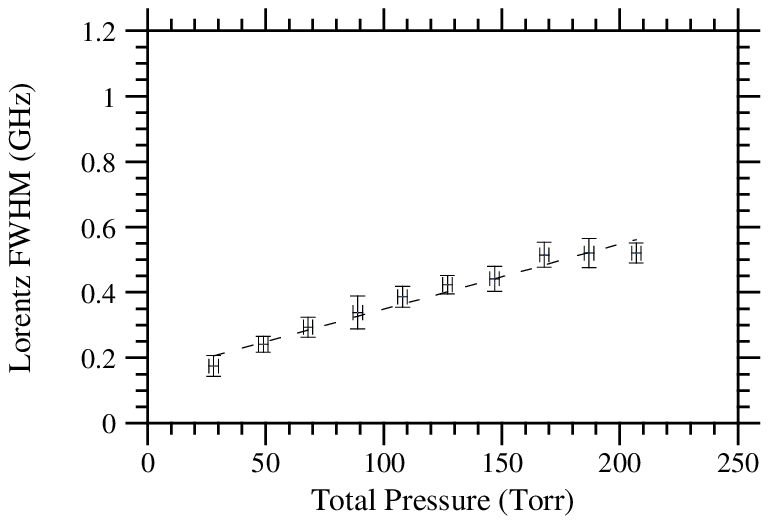}
}
\caption{FWHM He-broadening measurements for the 12713.30 cm$^{-1}$ methane absorption line as a function of pressure at room temperature. P$_\mathrm{CH_4}$ = 28 Torr.}
\label{ch4he}
\end{figure}

Table~\ref{tab:ch4b} shows the measured CH$_4$ pressure broadening coefficients by it-self, air, H$_2$ and He gases, at RT and only for a few more intense and well isolated lines, where the modulation amplitude could be kept low to continue using the second derivative approximation explained in Section~\ref{freq}. The shown errors are the maximum errors (3$\sigma$). The pressure line-shifting effect is usually about an order of magnitude lower then the broadening, but the sensitivity of this spectrometer does not permit to obtain reliable measurements, therefore they are not reported here.
\begin{table}
\caption{Methane pressure broadening FWHM coefficients ($\gamma$) at room temperature.}
\smallskip
\label{tab:ch4b}
\begin{tabular*}{.48\textwidth}{@{} c @{\extracolsep{\fill}} c @{\extracolsep{\fill}} c @{\extracolsep{\fill}} c @{\extracolsep{\fill}} c @{}}
\hline\noalign{\smallskip}
$\nu'$ & $\gamma_\mathrm{self}$ & $\gamma_\mathrm{air}$ & $\gamma_\mathrm{H_2}$ & $\gamma_\mathrm{He}$ \\
(cm$^{-1}$) & (MHz/Torr) & (MHz/Torr) & (MHz/Torr) &  (MHz/Torr) \\
\noalign{\smallskip}\hline\noalign{\smallskip}
12806.52 & $6.2 \pm 0.1$ & $5.0 \pm 0.1$ & $4.0 \pm 0.3$ & $4.5 \pm 0.4$ \\
12804.95 & $7.1 \pm 0.3$ & & & \\
12773.83 & $4.4 \pm 0.1$ & & & \\
12767.14 & $3.7 \pm 0.2$ & & & \\
12713.30 & $5.3 \pm 0.1$ & $4.8 \pm 0.4$ & $2.1 \pm 0.2$ & $2.0 \pm 0.2$ \\
\noalign{\smallskip}
\hline
\end{tabular*}
\end{table}
Examples of pressure broadening measurements are shown in Fig.~\ref{ch4self} and Fig.~\ref{ch4he}, where CH$_4$ broadening behaviors by collision with it-self and helium buffer gas respectively are displayed with their best linear fit for the 12713.30 cm$^{-1}$ line.

\noindent For the 780 nm band there are no known previous pressure broadening measurements, therefore a comparison can be done only with the results on different wavelengths. Moreover, due to the lack of quantum assignment of the lines in this band, only an average comparison can be done. Our broadening coefficients at RT in average are similar or perhaps a little lower than ours previous ones obtained at 840 nm~\cite{ref:alex} or to the ones coming from previous works on fundamentals, overtones or combination of them~\cite{ref:var,ref:vara,ref:fink,ref:fox,ref:rins,ref:varan,ref:zen,ref:gha,ref:pred,ref:monde,ref:men,ref:bou,ref:lyu,ref:ben,ref:hong} at RT. In the literature the self-broadening coefficients got from the ro-vibrational bands in the NIR and MID IR span from 6 to 7 MHz/Torr~\cite{ref:var,ref:bou,ref:men,ref:lyu}, while air-broadenings go from 4 to 6 MHz/Torr~\cite{ref:vara,ref:rins,ref:ben,ref:zen,ref:hong}. The H$_2$-broadening coefficients are between 5 MHz/Torr~\cite{ref:varan} and 6 MHz/Torr~\cite{ref:var}. The He-broadening coefficients got from the literature~\cite{ref:varan,ref:zen} for the $\nu_4$ and the $2\nu_2$ bands are in average $\simeq 3.1 - 3.3$ MHz/Torr.
From the HITRAN molecular spectroscopic database~\cite{ref:hit} in the range $11200 - 11300$ cm$^{-1}$ we found (4 $\leqslant \gamma_\mathrm{air} \leqslant 5.5$) MHz/Torr and (5.5 $\leqslant \gamma_\mathrm{self} \leqslant 6$) MHz/Torr,
The self-broadening, H$_2$-broadening and He-broadening coefficients in the visible~\cite{ref:keff}, at combination overtones higher than ours, give in average 7, 6 and 4 MHz/Torr respectively.
In any case as the broadening coefficients $\gamma$ are very dependent on the rotational quantum number $J$ ($\gamma$ values decrease as $J$ increases), these results are purely indicative until their quantum attribution will be clear.

\section{Conclusion}
\label{conc}
By using a tunable diode laser spectrometer with high resolving power ($\lambda/\Delta\lambda \approx 10^7$), the aid of the wavelength modulation spectroscopy technique with the second harmonic detection and a 30 m total path-length multipass measurement cell, 112 $^{12}$CH$_4$ absorption lines have been detected around 12780 cm$^{-1}$ and their positions measured within 0.01 cm$^{-1}$. They presumably belong to the combination overtone $3\nu_1 + \nu_3 + (\nu_2$ or $\nu_4)$ or $2\nu_1 + \nu_2 + 2\nu_3$ ro-vibrational bands. The line positions have been obtained by the comparison with reference I$_2$ absorptions  and the utilization of a very precise atlas. The maximum absorption cross section of the observed lines ranges between $1 \times 10^{-25}$ and $1 \times 10^{-24}$ cm$^2$/molecule at room temperature. The collisional broadening coefficients for different perturbing gases have been measured at RT for some of the more intense lines and a comparison with similar data from the literature has been tried.

\section{Acknowledgments}
The author is indebted to Davide Bertolini for the calculations in the high modulation approximation. Thanks are due to Alessandro Barbini for the electronic advising, Mauro Voliani for the mechanical assistance and Mauro Tagliaferri for the technical support.

\bibliography{CH4-ref.bib}

\end{document}